\documentclass[a4paper,12pt]{article}
\usepackage{graphicx,rotating,amsmath,amssymb}
\usepackage{hyperref}\usepackage{slashed,cite}
\pdfoutput=1

\hypersetup{colorlinks,bookmarksopen,bookmarksnumbered,
linkcolor=blus,pdfstartview=FitH,urlcolor=rossos,citecolor=verde}

\newcommand{\bp}{\bar{M}_{\rm Pl}}
\newcommand{\med}[1]{\langle #1 \rangle}

\newcommand{\fig}[1]{~\ref{fig:#1}}
\newcommand{\gs}{f_0}
\newcommand{\gt}{f_2}
\newcommand{\MS}{\overline{\mbox{\sc ms}}}

\usepackage{multicol}
\usepackage{color}
\definecolor{rosso}{cmyk}{0,1,1,0.4}
\definecolor{rossos}{cmyk}{0,1,1,0.55}
\definecolor{rossoc}{cmyk}{0,1,1,0.2}
\definecolor{blu}{cmyk}{1,1,0,0.3}
\definecolor{blus}{cmyk}{1,1,0,0.6}
\definecolor{bluc}{cmyk}{1,1,0,0.1}
\definecolor{verde}{cmyk}{0.92,0,0.59,0.25}
\definecolor{verdec}{cmyk}{0.92,0,0.59,0.15}
\definecolor{verdes}{cmyk}{0.92,0,0.59,0.4}
\newcommand{\mub}{\bar{\mu}}

\newcommand{\eq}[1]{~{\rm (\ref{eq:#1})}}

\newcommand{\diag}{\,{\rm diag}}
\def\circa#1{\,\raise.3ex\hbox{$#1$\kern-.75em\lower1ex\hbox{$\sim$}}\,}

\newcommand{\beq}{\begin{equation}}
\newcommand{\eeq}{\end{equation}}

\newcommand{\bea}{\begin{eqnarray}}
\newcommand{\eea}{\end{eqnarray}}
\newcommand{\be}{\begin{equation}}
\newcommand{\ee}{\end{equation}}
\font\tenrsfs=rsfs10 at 12pt
\font\sevenrsfs=rsfs7
\font\fiversfs=rsfs5
\newfam\rsfsfam
\textfont\rsfsfam=\tenrsfs
\scriptfont\rsfsfam=\sevenrsfs
\scriptscriptfont\rsfsfam=\fiversfs
\def\mathscr#1{{\fam\rsfsfam\relax#1}}
\def\Lag{\mathscr{L}}

\def\circa#1{\,\raise.3ex\hbox{$#1$\kern-.75em\lower1ex\hbox{$\sim$}}\,}
\makeatletter

\def\hhref#1{\href{http://arxiv.org/abs/#1}{arXiv:#1}} 
\def\baselinestretch{1.17}
\def\hhref#1{\href{http://arxiv.org/abs/#1}{arXiv:#1}}
\usepackage{xstring}
\newcommand{\hhrefq}[1]{\IfSubStr{#1}{:}{\href{http://inspirehep.net/search?ln=en&ln=en&p=#1&of=hb&action_search=Search&sf=&so=d&rm=&rg=25&sc=0}{InSpire:#1}}{\hhref{#1}}}

\def\art{\@ifnextchar[{\eart}{\oart}}
\def\eart[#1]#2#3#4#5#6{{\rm #2}, {\em #3 \bf #4} {\rm (#6) #5} ({\em #1})}
\def\article{\@ifnextchar[{\earticle}{\oarticle}}
\def\oarticle#1#2#3#4#5#6{{\rm #1}, {\ital `#6'}, {\rm #2 #3 (#5) #4}}
\def\earticle[#1]#2#3#4#5#6#7{{\rm #2}, {\ital `#7'}, {\rm #3 #4 (#6) #5}  [\hhrefq{#1}]}
\def\hepart[#1]#2{{\rm #2, \sl#1}}
\def\heparticle[#1]#2#3{#2, {\ital `#3'} [\hhrefq{#1}]}
\newcommand{\doi}[1]{\href{http://dx.doi.org/#1}{[link]}}
\newcommand{\hhrefqq}[1]{\IfBeginWith{#1}{10.}{\href{https://doi.org/#1}{doi:#1}}{\hhrefq{#1}}}

\renewenvironment{thebibliography}[1]
{\begin{multicols}{2}[\section*{\refname}]%
		\@mkboth{\MakeUppercase\refname}{\MakeUppercase\refname}%
		\list{\@biblabel{\@arabic\c@enumiv}}%
		{\settowidth\labelwidth{\@biblabel{#1}}%
			\leftmargin\labelwidth
			\advance\leftmargin\labelsep
			\@openbib@code
			\usecounter{enumiv}%
			\let\p@enumiv\@empty
			\renewcommand\theenumiv{\@arabic\c@enumiv}}%
		\sloppy
		\clubpenalty4000
		\@clubpenalty \clubpenalty
		\widowpenalty4000%
		\sfcode`\.\@m}
	{\renewcommand{\@noitemerr}
		{\@latex@warning{Empty `thebibliography' environment}}%
		\endlist\end{multicols}}

\font\ital=cmu10

\newcounter{alphaequation}[equation]
\def\thealphaequation{\theequation\hbox to
0.6em{\hfil\alph{alphaequation}\hfil}}
\def\eqnsystem#1{
\def\@eqnnum{{\rm (\thealphaequation)}}
\def\@@eqncr{\let\@tempa\relax \ifcase\@eqcnt \def\@tempa{& & &} \or
  \def\@tempa{& &}\or \def\@tempa{&}\fi\@tempa
  \if@eqnsw\@eqnnum\refstepcounter{alphaequation}\fi
\global\@eqnswtrue\global\@eqcnt=0\cr}
\refstepcounter{equation} \let\@currentlabel\theequation \def\@tempb{#1}
\ifx\@tempb\empty\else\label{#1}\fi
\refstepcounter{alphaequation}
\let\@currentlabel\thealphaequation
\global\@eqnswtrue\global\@eqcnt=0 \tabskip\@centering\let\\=\@eqncr
$$\halign to \displaywidth\bgroup \@eqnsel\hskip\@centering
$\displaystyle\tabskip\z@{##}$&\global\@eqcnt\@ne
\hskip2\arraycolsep\hfil${##}$\hfil& \global\@eqcnt\tw@\hskip2\arraycolsep
$\displaystyle\tabskip\z@{##}$\hfil
\tabskip\@centering&\llap{##}\tabskip\z@\cr}
\def\endeqnsystem{\@@eqncr\egroup$$\global\@ignoretrue} \makeatother

\begin{document}
\begingroup
\renewcommand{\baselinestretch}{1}
\begin{center}
{\LARGE \bf \color{rossos}
Infra-red enhanced loops in quadratic gravity}\\
\bigskip\bigskip
{\large\bf Alberto Salvio$^{a}$, Alessandro Strumia$^{b}$ {\rm and}  Marco Vitti$^{c}$}  \\[2ex]
{\it $^{a}$  Physics Department, University of Rome and INFN Tor Vergata, Italy}\\
{\it $^b$ Dipartimento di Fisica dell'Universit{\`a} di Pisa, Italia}\\
{\it $^c$ Karlsruhe Institute of Technology, Germany}

\bigskip\bigskip

{\large\bf\color{blus} Abstract}
\begin{quote}\large
It has been suggested that logarithmically enhanced infra-red loop corrections arising in theories with four derivatives 
correspond to a physical running of couplings, rendering  quadratic gravity asymptotically free. 
We find that these effects depend on the gauge and on the field parameterisation. 
We compute physical on-shell amplitudes and find genuine  infra-red log-enhanced loop corrections,
that are process-dependent and cannot be absorbed into running couplings. 
As a byproduct, we show that the effective action of quadratic gravity at tree level and without matter fields
is just Einstein gravity, so positivity bounds are satisfied despite the presence of ghosts.
\end{quote}
\thispagestyle{empty}
\end{center}
\tableofcontents
\endgroup
\setcounter{footnote}{0}
\newpage

\section{Introduction}
The authors of~\cite{2307.00055,2403.02397,2408.13142} claim
that theories with a dimension-less scalar
(such as an ordinary QFTs with 2 derivatives in 1+1 dimensions, or 4 derivative QFTs in $3+1$ dimensions)
contain IR divergences such that
the usual ultra-violet (UV) running (encoded in the dependence of renormalized
couplings on $\ln\mub$ in dimensional regularization)
does not match their dependence on $\ln p$, where $p$ denotes external energy-momenta. 

This issue affects quadratic gravity in 3+1 dimensions, which has possible physical interest
as a renormalizable theory of quantum gravity.
Indeed it contains, in the pure gravity sector, terms with 4 derivatives,
\beq \label{eq:Ladim}
S = \int d^4x \sqrt{|\det g|} \,\bigg[ \frac{R^2}{6\gs^2} + \frac{\frac13 R^2 -  R_{\mu\nu}^2}{\gt^2} -\frac{\bp^2}{2}R
+  \Lag_{\rm matter}
\bigg],
\eeq
where  $f_0$ and $f_2$ are two dimension-less couplings.\footnote{In~\cite{2403.02397} these parameters 
are denoted as $\xi = - 6f_0^2$ and $\lambda = f_2^2$.
Theories with 4 or higher derivatives contain a spin 2 ghost, which (so far) prevent making full sense of the theory when formulated initially in Minkowski rather than Euclidean space. The ghost issue is orthogonal to the issue addressed here, that also
affects 2-derivative theories  QFT in 1+1 dimensions.}
The second term is the square of the Weyl tensor, $ \frac13 R^2 -  R_{\mu\nu}^2=-W_{\alpha\beta\mu\nu}^2/2 $, 
and we can ignore the topological Gauss-Bonnet term.
The Einstein term induces masses $M_{0,2} = f_{0,2}\bp/\sqrt{2}$ for the spin-0 and spin-2 extra components
in the 4-derivative graviton.
The  one-loop $\beta$ functions, computed by regularising UV divergences
via dimensional regularization with scale $\mub$ at energies  larger than $M_{0,2}$, 
are~\cite{Avramidi:1985ki} (see also~\cite{Julve:1978xn,Fradkin:1981iu,1403.4226})
\beq\label{eq:RGG}
\frac{df_2^2}{d\ln\mub}= -\frac{133 \gt^4}{10 (4\pi)^2} ,\qquad
\frac{df_0^2}{d\ln\mub}=\frac{5 f_0^4  + 30 f_0^2 f_2^2 + 10 f_2^4 }{6(4\pi)^2}.
\eeq
The coupling $f_0$ is not asymptotically free, even including a generic matter sector~\cite{1403.4226}.

Usually, the running with the RG scale $\mub$ dictates the running with the typical external momentum $p$.
However, 4-derivative theories improve the UV convergence 
at the price of infra-red (IR) enhancements in tree-level~\cite{1808.07883} and loop~\cite{2403.02397} processes. 
The authors of \cite{2403.02397} interpret IR effects as a  running of couplings with external momenta $p$,
claiming a RG running that differs from the usual UV-running of eq.\eq{RGG}:
\beq\label{eq:RGE4dp}
\frac{df_2^2}{d\ln p} = -\frac{539 f_2^4+ 40 f_0^2f_2^2}{30(4\pi)^2 }, \qquad
\frac{df_0^2}{d\ln p} = \frac{f_0^4+6 f_0^2 f_2^2-70 f_2^4}{6(4\pi)^2}.
\eeq
These $\beta$ functions, computed in one simple gauge from the  one loop correction to the graviton propagator,
are interpreted as `physical' in~\cite{2403.02397} leading 
to the claim that quadratic gravity is asymptotically free~\cite{2403.02397}.
This interesting claim deserves a critical check.
We will show that:
\begin{enumerate}
\item Off-shell amplitudes (such as one-loop corrections to propagators) are not physical, 
as they change when using different parameterisations in field space.
In section~\ref{sec:freetoy} we illustrate how a simple {\bf field reparametrization generates an apparent $p$-running} even in a theory of a free scalar.
The issue of field parameterisations is unavoidable in gravitational computations:
 there is no preferred field basis where the action is a finite polynomial.
Any computation involves choosing an arbitrary basis in graviton field space,
such as the standard $g_{\mu\nu}(x)=\eta_{\mu\nu}+h_{\mu\nu}(x)$ expansion.

\item
In turn, this leads to {\bf gauge-dependence issues}.
In gauge theories (such as gravity) quantum effective actions appear to depend on the gauge fixing.
Closer inspection reveals that the gauge dependence is compensated by appropriate field reparametrizations~\cite{Nielsen},
meaning that different gauge fixings just select, at quantum level, different coordinates in field space.
We then expect that spurious gauge-dependent and $p$-dependent corrections can appear in unphysical quantities, such as one-loop off-shell propagators.
In section~\ref{sec:agravity} we re-compute the full quadratic gravity in more general gauges  than in~\cite{2403.02397},
finding that the one-loop graviton propagator is gauge dependent, including the terms interpreted as $p$-running in eq.\eq{RGE4dp}.
\end{enumerate}
In summary, we find that the `physical' $\ln p$ running of~\cite{2403.02397} is unphysical: 
it depends on how fields are parameterised, and on the gauge fixing chosen for reparametrization invariance.

The same computation confirms that the gauge-dependent off-shell graviton propagator can be used to extract the usual gauge-independent UV $\beta$ functions.
This is possible thanks to a special property of UV divergences: they only affect the parameters of the theory,
so that $\beta$ functions are gauge-independent.\footnote{We recall the standard proof (see e.g.~\cite{1307.3536}).
We expand a generic bare parameter $\theta_{\rm bare}$ measuring the
strength of a gauge invariant term in the Lagrangian 
as a perturbative series in terms of the 
renormalized parameter $\theta$.
The explicit expression  with gauge parameter $\xi$ is
\be 
\mub^{d-4}\theta_{\rm bare}= \theta + \frac{\theta_1 (\theta, \xi)}{d-4}+
\frac{\theta_2 (\theta, \xi)}{(d-4)^2}+ \cdots
\label{eq:lambda-definition}
\ee
in the $\MS$ scheme, where the $\theta_\ell$ for $\ell\ge 1 $ are defined to be the coefficients of the poles at $d=4$. 
Since $\theta_{\rm bare}$ is $\xi$-independent,
the renormalized $\theta$ and all $\ell$-loop $\theta_\ell$ must be gauge-independent too.
This implies  that the $\beta$ functions are gauge-independent.}
No analogous argument is expected to apply to IR divergences, as they arise from process-dependent long-range effects. 
Consequently, off-shell amplitudes can exhibit spurious IR runnings, even in free theories.

To establish how IR divergences affect quadratic gravity at loop level,
one needs to compute physical on-shell amplitudes, such as $g\to gg$ or $gg \to gg $, where $g$ is a gravitational particle.
We here compute one loop corrections in the simpler spin-0 sector, where there is no physical ghost.
This suffices to clarify the issue, because eq.\eq{RGE4dp} for $f_0$ differs from eq.\eq{RGG} even within the spin-0 sector.
In section~\ref{sec:0} we show that parameterising the graviton as $g_{\mu\nu}(x)=\eta_{\mu\nu} (1 + h/4)$ in terms of its trace $h$
reproduces the UV running in eq.\eq{RGG} and the anomalous IR running in eq.\eq{RGE4dp}.
However different parameterisations, such as  $g_{\mu\nu}(x)=e^{2\sigma(x)}\eta_{\mu\nu} $ and  $g_{\mu\nu}(x)=\Omega^2(x)\eta_{\mu\nu} $,
yield different IR runnings and the same UV running.
We also compute physical on-shell $2\to 2$ scatterings among spin 0 graviton components at one loop, 
and find that they are independent of the field parameterisation.
These amplitudes exhibit process-dependent  log-enhanced IR corrections, arising because the propagator has a stronger IR divergence than an ordinary propagator when the mass is taken to zero.
This behaviour contrasts with that of ordinary quantum field theories, 
where high-energy scattering processes are well-approximated by their tree-level expressions 
inserting couplings renormalized at the relevant high-energy scale. However, in quadratic gravity, scatterings cannot reach an energy so high~\cite{1808.07883} that these large process-dependent logarithms  invalidate perturbation theory.

To further clarify the issue, in section~\ref{sec:22} we decompose the 4-derivative field into two 2-derivative fields.
As well known the $R^2/6f_0^2$ term is equivalent to an extra scalar.
Since $f_0$ is just the scalar quartic in a power-counting renormalizable theory, 
it cannot contribute to the one-loop $\beta$ function of $f_2$, consistently with in eq.\eq{RGG} and in contrast with eq.\eq{RGE4dp}.
This second-order formulation reproduces the same physical amplitudes in a more transparent way, 
and shows that in theories with ghosts, seemingly super-renormalizable terms can influence high-energy physics, 
leading to nonstandard log-enhanced corrections.

In section~\ref{sec:concl} we conclude, summarising our understanding without discussing the difference with past works.

 \section{Toy example: a free scalar}\label{sec:freetoy}
We provide a simple example of unphysical $p$-running, by considering a {free} scalar $\tilde{s}(x)$
 with either 2 or 4 derivatives, described by the Lagrangian
 \beq 
\label{eq:Lag24} \Lag_{\tilde s} =\left\{ \begin{array}{ll}
[ (\partial_\mu \tilde{s})^2  - m^2 \tilde{s}^2]/2 & \hbox{i.e.\ $\Pi_0(p) = (p^2-m^2)$ }
\cr
 [(\partial^2 + m_1^2)\tilde s][(\partial^2 + m_2^2)\tilde s]/2 & \hbox{i.e.\ $\Pi_0(p)= (p^2-m_1^2)(p^2-m_2^2)$}
 \end{array}\right.
 \eeq
 so that its propagator is $D(p) ={i}/\Pi_0(p)$. 
We perform a local change of field variables, $\tilde s(x) = F(s(x))$, where $F$ is a real invertible function.
The functional integration measure $D\tilde{s}$ acquires a Jacobian determinant under the field redefinition~\cite{Chisholm:1961tha,Kamefuchi:1961sb,tHooft:1973wag,1205.3279,2105.11482,2312.06748}:
\be D\tilde{s}= \prod_x d\tilde{s}(x) \propto e^{\delta^{(d)}(0) \int d^d x \, \ln |F'(s(x))|} \prod_x ds(x),\qquad \delta^{(d)}(0) = \int \frac{d^d k}{(2\pi)^d} = 0, \label{InvM} \ee
where $F' = dF/ds$.
The extra field-dependent exponent vanishes in dimensional regularization because of the vanishing of the $d$-dimensional delta function evaluated at zero.
More in general, the Jacobian can be computed as a functional determinant by introducing an unphysical ghost 
with momentum-independent propagator~\cite{Marcus:1984ei}.
We employ dimensional regularization, and perform the leading-order field redefinition, invertible around small $s$,
\beq \label{eq:trans}
 \tilde s(x) = s(x) + g s^2(x).\eeq
The field reparametrization of eq.\eq{trans} introduces in $\Lag_s$
an apparent cubic interaction $gs^3$ (which hides the $\tilde{s}\to -\tilde{s}$ symmetry, but 
vanishes on-shell) and a quartic $g^2 s^4$ interaction.
The apparent coupling $g$ is dimension-less in the 4-derivative theory, and has dimension $1$/mass in the $2$-derivative theory.
Nevertheless physical amplitudes, both at tree and loop level, recognise that $s$ is a free scalar and vanish on-shell, 
while exhibiting apparent $p$-runnings off-shell.
For example, the quadratic effective action for $s$ at one loop level is, in both theories of eq.\eq{Lag24},\footnote{This can be obtained
from usual momentum-space Feynman diagrams,  summing the bubble, seagull and tadpole diagrams.
Its structure, and the general vanishing of on-shell corrections, can be understood from
a simpler coordinate space computation, by perturbatively inverting eq.\eq{trans} as
$ s(x) = \tilde s(x) - g \tilde s^2(x) + 2 g^2 \tilde s^3(x)- 5 g^3\tilde{s}^4(x)+\cdots$
to relate the interacting theory for $s$ to composite operators of the free $\tilde{s}$ theory.
The one-point function up to order $g$  corresponds to the one-loop tadpole:
\beq \med{s(x)} = - g \med{\tilde s^2(x)}+{\cal O}(g^3) = -g D(0) . \eeq
The two-point function up to $g^2$ order is 
\begin{eqnarray} \label{eq:medsxs0}
\med{s(x) s(0)} &=& \med{\tilde{s}(x) \tilde{s}(0)}+g^2  \bigg[\med{\tilde s^2(x)\tilde{s}^2(0)}+
2 \med{\tilde{s}(x) \tilde{s}^3(0)}+
2 \med{\tilde{s}^3(x) \tilde{s}(0)}\bigg]+{\cal O}(g^3)\\
&=& D(x) + g^2 [2D(x)^2 + D(0)^2 + 12 D(x) D(0)] +{\cal O}(g^3)\label{eq:medsxs0p}.
\end{eqnarray}
In Fourier transform this equation becomes 
\beq \label{eq:prop1loop}
 \frac{i}{\Pi_0(p)}  +2ig^2 \left[ B(p)+\frac{6A}{\Pi_0(p)}
 \right] +{\cal O}(g^3).
\eeq
Inverting eq.\eq{prop1loop} leads to eq.\eq{prop1s},
as the extra $A^2$ term has vanishing coefficient $\delta^{(d)}(p) =0$ in dimensional regularization.}
\beq \label{eq:prop1s}
\Pi(p) = \Pi_0(p) -2 g^2 \Pi_0(p) [6 A +\Pi_0(p) B(p) ].\eeq
Here $A$ and $B$ are loop functions that  reduce to the usual Passarino-Veltman functions $A_0$ and $B_0$, in the 2-derivative theory: 
 \beq A = \frac{1}{i}
 \int\frac{d^dk}{(2\pi)^d} D(k),\qquad
B(p)= \frac{1}{i} \int\frac{d^dk}{(2\pi)^d} D(k) D(p-k).\eeq
$A$ and $B$ induce apparent UV divergences, that just renormalise the field redefinition of eq.\eq{trans}, see e.g.~\cite{1605.03602}.
Furthermore,  $B(p)$  induces an apparent off-shell $p$-running.
This only arises at sub-leading order in large $p\gg m$ in the 2-derivative theory with apparent non-renormalizable interaction $g$:
\beq (p^2-m^2)^2 B(p) \stackrel{p\gg m}{\simeq} \frac{p^4 }{(4\pi)^2} \bigg[\frac{1}{\epsilon}+\ln\frac{\mub^2}{p^2}\bigg] -2  \frac{m^2 p^2  }{(4\pi)^2} \left[\frac{1}{\epsilon}+
\ln\frac{m^2\mub^2}{p^4}\right] + {\cal O}(m^4).\eeq 
For simplicity, we considered $m_1 = m_2 = m$.
In the 4-derivative theory the apparent  interaction $g$ is renormalizable and an apparent $p$-running arises at leading order in $p \gg m$
\beq (p^2-m^2)^4 B(p) \stackrel{p\gg m}{\simeq} \frac{2p^4}{(4\pi)^2} \ln\frac{p^2}{m^2}+{\cal O}(m^2 p^2, m^4).\eeq
Denoting the two on-shell states by their masses $m_{1,2}$,
we verified the expected vanishing of the one loop on-shell $m_2 \to m_1 + m_1$ decay amplitude,
and of the $m_1 + m_1\to m_1 + m_1$ scattering amplitude
(after combining 1PI corrections to $2,3,4$ vertices, including tadpole diagrams).

As a side remark, we note that applying the functional renormalisation group to these free theories 
would generate apparent interactions, due to the presence of a finite momentum cutoff and the use of approximate truncations.

\section{Quadratic gravity}\label{sec:agravity}
We consider the full quadratic gravity of eq.\eq{Ladim}, which we
quantise expanding around the flat-space Minkowski metric $\eta_{\mu\nu}=\diag(1,-1,-1,-1)$ as $g_{\mu\nu} = \eta_{\mu\nu} + h_{\mu\nu}$.
The background solves the classical equations provided that the vacuum energy vanishes
at quantum level, thanks to unspecified interactions in $\Lag_{\rm matter} = - V + \cdots$
that cancel the graviton loop correction to the graviton tadpole, proportional to $M_0^4+5 M_2^4$.
The simplest possibility is a constant $V$ counter-term.
For this reason we omit the graviton tadpole in the following computations.

Although standard, the fluctuation $h_{\mu\nu}$ represents just 
one arbitrary field parameterisation: for example, $h^{\mu\nu}$ would be a different parameterisation.
Based on our previous discussion, we expect that gauge fixing re-opens the  issue of field re-parameterisations at loop level. 
To check this, we expand the computation of~\cite{2403.02397} considering several different gauge fixings.
We will find that the usual UV $\beta$ function remains gauge-independent, whereas the IR $\beta$ function of~\cite{2403.02397} is gauge-dependent.

\subsection{Gauge-fixing of quadratic gravity}\label{sec:gf}
Gravity is gauge invariant under coordinate reparametrizations: 
we quantise eq.\eq{Ladim} following the Fadeev-Popov procedure with gauge condition 
\beq \label{eq:gfc}
f_\mu = \partial_\nu\bigg( h_{\mu\nu} -\frac{ c_g }{2} \eta_{\mu\nu} h_{\alpha \alpha}\bigg)
\eeq
where $c_g$ is a free gauge parameter.
Gauge-invariant quantities can be computed integrating over an arbitrary smearing of
the gauge condition $f_\mu=0$~\cite{Stelle:1976gc,Avramidi:1985ki,hep-th/0412249,1308.3398}.
We use this freedom to choose a gauge fixing term containing two extra free gauge parameters $\xi_g$ and $d_g$
and extra derivatives:
\beq \label{eq:gf}
S_{\rm gf} = - \frac{1}{2\xi_g}\int d^4x ~ f_\mu Y_{\mu\nu} f_\nu,\qquad
Y_{\mu\nu}=\partial^2 \eta_{\mu\nu} + (d_g-1) \partial_\mu \partial_\nu .
\eeq
We choose a non-covariant gauge fixing purely quadratic in $h_{\mu\nu}$, such that gauge fixing does not affect the graviton couplings,
and such that $\det Y$ is a field-independent constant.
The ghost action is not affected by $\xi_g$ nor $d_g$ and remains the same as in~\cite{1403.4226}, where $d_g=1$.

At quadratic level the purely gravitational action becomes
\begin{eqnarray} 
S &=& \frac12 \int d^4k~ {h}_{\mu\nu}  \bigg[
 \bigg(-\frac{k^4}{2f_2^2 } +\frac{\bp^2 k^2}{4}\bigg) P^{(2)}
+ 
\bigg(\frac{k^4}{\gs^2}+\frac{3c_g^2 d_g k^4}{4\xi_g} -\frac{\bp^2 k^2}{2} \bigg) P^{(0s)}  +
 \nonumber \\
&&
+ \frac{k^4}{2\xi_g} P^{(1)}
+ k^4\frac{(c_g-2)^2 d_g}{4\xi_g}{P}^{(0w)} 
+k^4 \frac{\sqrt{3}c_g(c_g-2)d_g}{4\xi_g} T^{(0)} \bigg]_{\mu\nu\rho\sigma}   h_{\rho\sigma}\label{eq:Sg2}
\end{eqnarray}
where  
\begin{eqnsystem}{sys:P}
P^{(2)}_{\mu\nu\rho\sigma} &=& \frac12 T_{\mu\rho}T_{\nu\sigma} + \frac12 T_{\mu\sigma} T_{\nu\rho}-\frac{T_{\mu\nu}T_{\rho \sigma}}{d-1} \\
P^{(1)}_{\mu\nu\rho\sigma} &=& \frac12 (T_{\mu\rho}L_{\nu\sigma} +  T_{\mu\sigma} L_{\nu\rho}+T_{\nu\rho}L_{\mu\sigma} +  T_{\nu\sigma} L_{\mu\rho})\\
P^{(0s)}_{\mu\nu\rho\sigma} &=& \frac{T_{\mu\nu}T_{\rho \sigma}}{d-1} \\
 P^{(0w)}_{\mu\nu\rho\sigma} &=& L_{\mu\nu}L_{\rho \sigma}
 \end{eqnsystem}
are projectors over spin-2, spin-1 and spin-0 components of $h_{\mu\nu}$, written in terms of $d=4-2\epsilon$,
$T_{\mu\nu} = \eta_{\mu\nu} - k_\mu k_\nu/k^2$ and $ L_{\mu\nu}=k_\mu k_\nu/k^2$.
Their sum equals unity: 
\beq 
(P^{(2)}+P^{(1)}+P^{(0s)}+P^{(0w)})_{\mu\nu\rho\sigma}=\frac12 (\eta_{\mu\rho}\eta_{\nu\sigma}+\eta_{\mu\sigma}\eta_{\rho\nu}).\eeq
Furthermore 
$T^{(0)}  _{\mu\nu\rho\sigma} =P^{(0sw)}_{\mu\nu\rho\sigma} + P^{(0ws)}_{\mu\nu\rho\sigma}$, with 
 $P^{(0sw)}_{\mu\nu\rho\sigma} =T_{\mu\nu}L_{\rho\sigma}/\sqrt{d-1}$ and $P^{(0ws)}_{\mu\nu\rho\sigma} =  L_{\mu\nu}T_{\rho\sigma}/\sqrt{d-1}$.

\smallskip

The graviton propagator is obtained inverting the terms quadratic in $h_{\mu\nu}$. Let us consider a generic gauge, where these quadratic terms read
\begin{eqnarray} 
S &=& \frac12 \int d^dk~ {h}_{\mu\nu}  \bigg[ c_2 P^{(2)}_{\mu\nu\rho\sigma} + c_1P^{(1)}_{\mu\nu\rho\sigma} 
+ c_s P^{(0s)}_{\mu\nu\rho\sigma}   +c_w{P}^{(0w)}_{\mu\nu\rho\sigma} + c_{sw}{P}^{(0sw)}_{\mu\nu\rho\sigma}+ c_{ws}{P}^{(0ws)}_{\mu\nu\rho\sigma}\nonumber \bigg]  h_{\rho\sigma},\label{genQuad}
\end{eqnarray}
where $c_2, c_1, c_s, c_w, c_{sw}, c_{ws}$ are functions of $k^2$.
Inverting the quadratic action gives the graviton propagator
\beq\label{genericD}
D_{\mu\nu\,\rho\sigma} =
i \bigg[ \frac{P^{(2)}}{c_2} +\frac{P^{(1)}}{c_1} +\frac{c_wP^{(0s)}+c_sP^{(0w)}-c_{sw}P^{(0sw)}-c_{ws}P^{(0ws)}}{c_s c_w-c_{sw} c_{ws}}
  \bigg]_{\mu\nu\rho\sigma} .
\eeq
Specialising this formula to eq.~(\ref{eq:Sg2}) gives
\beq\label{eq:gravprop}\small
\hspace{-0.5cm}D_{\mu\nu\,\rho\sigma} =
\frac{i}{k^2} \bigg[ \frac{- 2 \gt^2 P^{(2)} }{k^2 - M_2^2} +
\frac{2\xi_g  P^{(1)} }{k^2}+ 
\frac{\gs^2  }{ k^2- M_0^2}\bigg(P^{(0s)}+\frac{\sqrt{3}  c_g T^{(0)}}{2-c_g}  \bigg)+
\frac{e_g P^{(0w)}}{k^2-M_0^2}-
\frac{4M_0^2 \xi_g P^{(0w)} }{(c_g-2)^2 d_g k^2 (k^2-M_0^2)}
  \bigg]_{\mu\nu\rho\sigma} 
\eeq
where $e_g= (3 c_g^2 f_0^2 + 4\xi_g/d_g)/{(2-c_g)^2}$ and 
$M_{0,2}^2 = \bp^2 f_{0,2}^2/2$ are the masses of the spin-0 graviton and of the spin-2 ghost.
Ignoring these masses by setting $\bp=0$, the graviton propagator simplifies to
\beq\label{eq:gravprop2}
D_{\mu\nu\,\rho\sigma} =
\frac{i}{k^4} \bigg[ - 2 \gt^2 P^{(2)}+  f_0^2 \bigg(P^{(0s)}+\frac{\sqrt{3}  c_g T^{(0)}}{2-c_g}  \bigg)
+ 2\xi_g  P^{(1)} + e_g P^{(0w)}
  \bigg]_{\mu\nu\rho\sigma} .
\eeq
In this limit $M_{0,2}=0$ all longitudinal terms cancel giving the
simple Feynman-like form 
 \beq\label{eq:PercacciGauge}
D_{\mu\nu\,\rho\sigma} =
\frac{i}{k^4} \bigg[ -f_2^2 (\eta_{\mu\sigma}\eta_{\nu\rho}+\eta_{\mu\rho}\eta_{\nu\sigma})+
\frac{2f_2^2+f_0^2}{d-1}\eta_{\mu\nu}\eta_{\rho\sigma}\bigg]
\eeq
by choosing the gauge-fixing parameters as
\beq \xi_g=-f_2^2,\qquad
c_g =2\frac{f_0^2 + 2f_2^2}{df_0^2+2f_2^2},\qquad 
e_g= \frac{f_0^2-2(d-2) f_2^2}{d-1}.
\eeq
This propagator agrees with the one used in~\cite{1308.3398,2403.02397}, where however the gauge fixing $S_{\rm gf}$ was chosen to
be covariant: so our gauge differs from the gauge used in~\cite{1308.3398,2403.02397}, that contains
extra graviton interactions and a consequent `third ghost'~\cite{Avramidi:1985ki,hep-th/0412249,1308.3398}.

\begin{figure}[t]
$$\includegraphics[width=\textwidth]{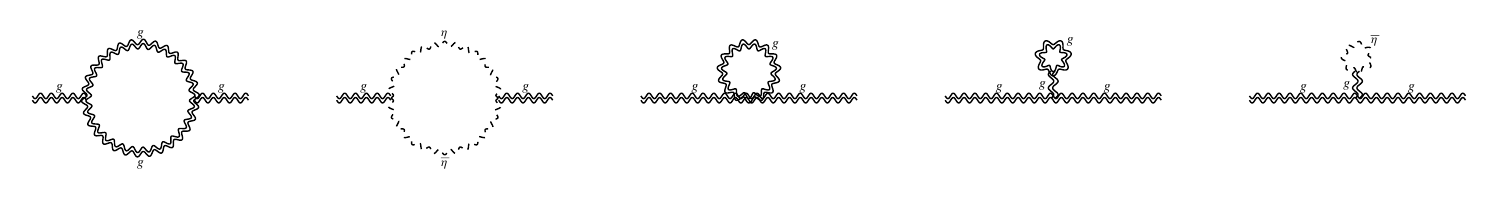}$$
\begin{center}
\caption{\em\label{fig:Feyns} One loop gravitational correction the graviton propagator. 
The tadpole diagrams and the ghosts $\eta$ must be included in the 1PI action.}
\end{center}
\end{figure}

\subsection{The graviton propagator at one loop}
We compute the one-loop correction to the graviton propagator using Feynman-diagram techniques.
In order to test the gauge (in)variance of the result we employ the more general gauge presented in section~\ref{sec:gf},
resulting in the graviton propagator of eq.\eq{gravprop}.
It depends on three gauge parameters $\xi_g, c_g,e_g$.

First, we reproduce the usual $\beta$ functions.
We employ dimensional regularization in $d=4-2\epsilon$ dimensions with scale $\bar\mu$.
We keep the masses $M_{0,2}$ to regularise the IR divergences of the $1/k^4$ propagator in the massless limit.
By computing the one-loop correction to the interaction of one 4-derivative graviton with
a generic particle (for simplicity we employ a scalar with momentum $p$), we find
the correct tensorial structure $\frac12 p^2 \eta_{\mu\nu} - p_\mu p_\nu$, provided that the 
graviton field is renormalized as follows:
\beq    
h_{\mu\nu} \to \frac{1}{\sqrt{Z_{TL}}} \bigg(h_{\mu\nu}-\frac{1}{4}\eta_{\mu\nu} h_{\alpha\alpha}\bigg)+
\frac{1}{\sqrt{Z_{T}}} \frac{1}{4}\eta_{\mu\nu} h_{\alpha\alpha}.
\label{eq:Zg}\eeq
The wave-function renormalization $Z_T$ and $Z_{TL}$ differ 
because we used a simple gravitational gauge-fixing term that breaks general relativity but respects special
relativity: thereby distinct representations of the Lorentz group
(the trace and the traceless part of $h_{\mu\nu}$), 
get different renormalizations (see Refs.~\cite{1403.4226,2305.10591} for previous discussions of wave-function renormalization in quadratic gravity).
We find the one-loop results 
\bea
Z_T &=& 1 + \frac{1}{(4\pi)^2\epsilon} \left[\frac{3 c_g^2}{4(c_g-2)^2} f_0^2 - \frac{e_g}{4} + 3 \xi_g\right], \\
Z_{TL}&=& 1 + \frac{1}{(4\pi)^2\epsilon}\bigg[
\frac{10}{9}\frac{c_g-4}{c_g-2} f_2^2+
\frac{c_g^2+12c_g-16}{18(c_g-2)^2} f_0^2 +\frac43\xi_g -\frac{e_g}{6}
\bigg].
\eea
We next compute the one-loop correction to the graviton propagator,
described by the Feynman graphs in fig.\fig{Feyns}.
We extend the similar computation in~\cite{1403.4226} by using a more general gauge and by 
computing also the finite terms, in particular the IR-enhanced finite  $\ln p$ terms at $p\gg M_{0,2}$.
After applying the graviton field renormalisation of eq.\eq{Zg}, 
the tensorial structure of the one-loop graviton kinetic term matches the structure 
of the action in eq.\eq{Sg2}, allowing to isolate the corrections to its spin-2 and spin-0 part as
\beq \sum_{i = \{0,2\}} 
\left\{
k^4 P^{(i)}_{\mu\nu\rho\sigma}
\left[ c^{\rm UV}_i \left(\frac{1}{\epsilon} + \ln\frac{\bar\mu^2}{M_{0,2}^2}\right)
+ c_i^{\rm IR} \ln \frac{M_{0,2}^2}{k^2} \right] \right\}+
\hbox{finite and longitudinal terms}.
\label{eq:resUV}
\eeq
The coefficients $c^{\rm UV}_{0,2}$ reproduce the usual gauge-independent one-loop $\beta$ functions of eq.~(\ref{eq:RGG}).
The terms $c^{\rm IR}_{0,2}$ are found to be gauge-dependent, and different from the ones in eq.\eq{RGE4dp}, computed in one different gauge.
We do not report their complicated expressions.

\subsection{The effective action of quadratic gravity}
While we cannot compute on-shell one-loop $2\to 2$ scattering amplitudes among 
components of the 4-derivative graviton multiplet,
an 
 aside feature arises at tree level:
the scattering amplitudes for the massless graviton mode (selected by setting $p_i^2=0$ in off-shell amplitudes) are the same as in Einstein gravity.
This result was previously observed and understood  by noting that the quadratic-gravity classical action can be obtained from the Einstein-Hilbert classical action by substituting 
$g_{\mu\nu}$ with
\beq \label{eq:Anselmi}
g_{\mu\nu}'= g_{\mu\nu}- \frac{g_{\mu\nu}R}{6 f_0^2} 
+ \frac{ \frac16 g_{\mu\nu}R-R_{\mu\nu}}{f_2^2}+ {\cal O}(f_0,f_2)^{-4}.\eeq
This implies that the Effective Field Theory (EFT) obtained from pure quadratic gravity
integrating out at tree level the massive spin 0 and spin 2 fields is Einstein gravity.
However, quadratic gravity and Einstein gravity are not equivalent as eq.~(\ref{eq:Anselmi})
does not bi-univocally cover the full field space: the transformed action only contains the massless graviton,
and the extra modes of quadratic gravity are expected to produce loop-level corrections~\cite{Marcus:1984ei,hep-th/0605205,1506.04589}.

The tree-level effective action fulfils general EFT positivity bounds on the signs of coefficients of
non-renormalizable operators~\cite{2103.12728,2201.06602} (see also~\cite{2304.02550}), despite that the full theory contains a ghost.\footnote{This result might seem trivial, since a field redefinition as in eq.\eq{Anselmi} allows to 
remove quadratic terms from a generic gravitational theory~\cite{Marcus:1984ei}.
In general, this removal is achieved at the price of generating an infinite tower of higher-dimensional operators. 
In quadratic gravity such operators are assumed to be absent to get renormalizability.
So the result that its EFT at tree level is just the Einstein-Hilbert action (with no quadratic, nor cubic, nor any other term) is physically meaningful.}

In the next section we compute the loop-level amplitude in the spin 0 sector.

\section{The spin 0 sector of quadratic gravity}\label{sec:0}
In this section we study the spin 0 sector (characterized by the $f_0$ coupling) in the limit $f_2\ll f_0$,
as this can be done by restricting the graviton $g_{\mu\nu}$ to its ghost-free spin-0 trace and ignoring the
more complicated spin-2 part~\cite{1705.03896}.
The perturbative path-integral measure of gravity can be splitted by singling out the graviton trace as~\cite{Salvio:2024joi}
\be \prod_x \bigg( \prod_{\mu\leq\nu} dh_{\mu\nu}(x)\bigg) \propto \prod_x \bigg(dh(x) \prod_{\mu\leq\nu} dh^{\rm TL}_{\mu\nu}(x)\bigg), \label{grM}\ee
where the first product is over all spacetime points $x$ and we decomposed  $h_{\mu\nu}= h^{\rm TL}_{\mu\nu}+\eta_{\mu\nu} h/d$,
with $h^{\rm TL}_{\mu\nu}$ being the traceless part of $h_{\mu\nu}$. In eq.~(\ref{grM}) we ignored field-independent overall constants that are irrelevant for our purposes. 
This means that we can simply restrict the graviton $g_{\mu\nu}$ to be proportional to $\eta_{\mu\nu}$, such that the Weyl term vanishes.
The proportionality factor is a scalar, that can be written in different basis in field space.
We will examine  the following choices in terms of $h$ or $\sigma$ or $\Omega$:
\beq \label{eq:g2trace}
g_{\mu\nu}(x)=\left [1+\frac{h(x)}{d}\right]\eta_{\mu\nu}   \equiv e^{2\sigma(x)} \eta_{\mu\nu} \equiv  \Omega^2(x) \eta_{\mu\nu}.\eeq
In the small-field limit these field transformations reduce to eq.\eq{trans} with specific values of $g$.

\subsection{Spin-0 sector in terms of the graviton trace $h$}\label{sec:h}
We first compute using the standard graviton field parameterisation in eq.\eq{g2trace}, such that
\beq \sqrt{|\det g|} = (1+h/4)^2,\qquad  R = - 6(1+h/4)^{-3/2}   \partial^2 \sqrt{1+h/4}. \eeq
The action of eq.\eq{Ladim} can then be perturbatively  expanded for $h\ll 1$, obtaining an infinite series
that obscures its renormalizability
\begin{eqnsystem}{sys:Lagh} \label{eq:Lagh}
\Lag_h  &=& \frac{3}{16 f_0^2}\left[ \frac{[(\partial h)^2/2 - 4(1+h/4) \partial^2 h]^2}{32(1+h/4)^4}-
M_0^2 \frac{(\partial h)^2/2 - 4(1+h/4) \partial^2 h}{1+h/4}\right]- V\bigg(1+\frac{h}{4}\bigg)^2\qquad \\
&=&\frac{3}{16 f_0^2} \left[\frac{ (\partial^2 h)^2- M_0^2 (\partial h)^2}{2}-
 \frac{2h (\partial^2 h)^2+ (\partial h)^2 (\partial^2 h) - M_0^2 h (\partial h)^2}{8}  + \cdots\right].
\end{eqnsystem}
This action shows that $h$ is a massless ghost in the Einstein limit.\footnote{In case this appears confusing, it can be understood as follows.
In the full gravitational theory the graviton trace is unphysical because 
coordinate reparametrizations $x_\mu\to x_\mu+\xi_\mu(x)$ allow to remove it.
This is no longer true when restricting the graviton to its trace imposing eq.\eq{g2trace} 
because this ansatz restricts the allowed $\xi_\mu(x)$ to be conformal transformations, which include Poincar\'e,
scale and special conformal transformations (see e.g.~\cite{1705.03896}).
Since actions for the graviton trace are only invariant under conformal transformations,
the graviton trace has a well-defined propagator and no extra gauge
fixing of reparametrizations is needed.
Conformal transformations act on $\sigma$ as
\beq \label{eq:conformal}
\sigma(x)\to \sigma(x')-\frac{d}2 c\cdot x, \qquad x'_\mu=x_\mu+ c_\mu x^2 - 2 x_\mu (c\cdot x).\eeq
Scale transformations act as $x'=(1+\lambda) x$ and $\sigma(x)\to \sigma(x')- \lambda$.
}
Adding the $R^2/6 f_0^2$ term, $h$ becomes a 4-derivative field that also contains a physical spin-0 scalar with mass $M_0 = f_0 \bp/\sqrt{2}$.
Eq.\eq{Lagh} reproduces the graviton interactions in the full computation, where the graviton expanded as
$g_{\mu\nu}=\eta_{\mu\nu}+h_{\mu\nu}$ is next projected over its trace. 
The two degrees of freedom in the 4-derivative field can be selected, when computing amplitudes, by
putting the external momenta $p_i$ of each particle $i$ either at $p_i^2=0$ or at $p_i^2 = M_0^2$.
Tree-level amplitudes vanish non-trivially on-shell at $p^2_i=0$, namely for scatterings involving the massless mode.
The $2\to 2$ scattering amplitudes are
{\small\begin{eqnsystem}{sys:Amplitudes}
\mathscr{A}_{0 0 \to 0 0}&=&\mathscr{A}_{0 0 \to 0 M_0}=0\\
\frac{\mathscr{A}_{0 0 \to M_0 M_0}}{N}&=&\frac{ i f_0^2}{6} \left(1+\frac{2M_0^2}{t-M_0^2}+\frac{2 M_0^2}{u-M_0^2}\right)\stackrel{s,t\gg M_0^2}\simeq \frac{i f_0^2 }{6}\\
\frac{\mathscr{A}_{0 M_0 \to M_0 M_0}}{N}&=&\frac{ if_0^2}{2}   \left(1+\frac{2M_0^2}{s-M_0^2}+\frac{2 M_0^2}{t-M_0^2}+\frac{2 M_0^2}{u-M_0^2}\right)
\stackrel{s,t\gg M_0^2}\simeq \frac{i f_0^2 }{2}\\
\frac{\mathscr{A}_{M_0 M_0 \to M_0 M_0}}{N} &=&
i f_0^2 \left(1+\frac{3 M_0^2 }{s-M_0^2}+\frac{3 M_0^2}{t-M_0^2}+\frac{3M_0^2 }{u-M_0^2}-\frac{M_0^2}{3s}-\frac{M_0^2}{3t}-\frac{M_0^2}{3u} \right)
\stackrel{s,t\gg M_0^2}\simeq i f_0^2 \qquad
\end{eqnsystem}}%
where $N=(\sqrt{16/3f_0} M_0)^4$ makes amplitudes canonically normalised (possibly up to a sign).
Individual diagrams contribute with terms that grow with higher powers of energy,
and the simple structure of eq.~(\ref{sys:Amplitudes}) emerges only after cancellations.
This suggests that the $h$ basis introduces artificial off-shell complications, similar to those encountered in unitary gauges.

The tree level 2-particle effective action 
for $h$ is $\Pi_0^h (p) = 3 p^2(p^2-M_0^2)/16f_0^2$. 
The one loop correction in the limit $M_0=0$, arising from the bubble and seagull diagrams 
 (assuming, as discussed above, that additional terms cancel the graviton tadpole contribution to the vacuum energy), is
\beq {i}\Pi_1^h (p) = \int\frac{d^dk}{(2\pi)^d}\left[-\frac18 \frac{[k^2+p^2+k\cdot p]^4}{k^4(k+p)^4}+
\frac{6k^4+6p^4+13k^2p^2+2 (p\cdot k)^2}{32k^4}\right].
\eeq
Both diagrams are IR divergent and UV divergent.
Restoring the $M_0$ mass to control IR divergences
and computing in the off-shell limit $p\gg M_0$ gives
\beq \frac{\Pi_1^h(p)}{\Pi_0^h(p) } \simeq \frac{5f_0^2  }{12(4\pi)^2} \left[\frac{1}{\epsilon} +\ln \frac{-\mub^2}{M_0^{8/5}p^{2/5}} +3\right].
\label{eq:Pi1h}
\eeq
The unusual logarithmic term arises because, for example, the external momentum $p$ does not circulate in the loop in
the seagull diagram (hence no $\ln p$ dependence from that diagram ), yet it contributes a term growing as $p^4$ at large off-shell $p\gg M_0$.
In the limit $M_0 ,V\to 0$ all UV divergences are reabsorbed by a renormalisation of $f_0$ 
without requiring a wave-function renormalisation for the non-canonical field $h$:
\beq \label{eq:f0renorm}
f_0^2 \to f_0^2 \left[1+ \frac{5 f_0^2}{12\epsilon(4\pi)^2}\right], \qquad h\to h.\eeq
Indeed, all terms in $\Lag_h$ of eq.\eq{Lagh} with different powers of $h$
have coefficients proportional to $1/f_0^2$ and receive a UV-divergent one-loop correction
equal to the tree level term times the common loop/tree ratio in eq.\eq{Pi1h}.
The resulting beta function is $\beta_{f_0^2} =5 f_0^4/6(4\pi)^2$ in agreement with eq.\eq{RGG}.
The $\ln p$ term in eq.\eq{Pi1h} has a coefficient which agrees with eq.\eq{RGE4dp}~\cite{2403.02397} and which 
differs from the coefficient of the $1/\epsilon$ pole.

\subsection{Spin-0 sector in terms of the conformal $\sigma$}\label{sec:sigma}
Next, we compute using the second field parameterisation in eq.\eq{g2trace}
for the spin 0 graviton mode, $g_{\mu\nu}(x) = e^{2\sigma(x)} \eta_{\mu\nu}$. 
This leads to $R = - 6 e^{-3\sigma} \partial^2 e^\sigma$ and to the action
\beq \Lag_\sigma = \frac{6}{f_0^2}\bigg[ \big(\partial^2 \sigma + (\partial \sigma)^2\big)^2 -
M_0^2 e^{2\sigma} (\partial\sigma)^2\bigg]
-V \,  e^{4\sigma}. \label{eq:LagSM}
\eeq
On-shell tree amplitudes are the same as in the previous basis, eq.~(\ref{sys:Amplitudes}).

The two-particle action at tree level is $\Pi_0^\sigma (p)=  12 p^2(p^2-M_0^2)/f_0^2$. 
When dimensionfull terms $M_0, V$ vanish,
$\Lag_\sigma $ is a 4-th order polynomial in $\sigma$, invariant under conformal transformations~\cite{1705.03896},
and the one loop correction
to the two-point function is the sum of a bubble and a seagull diagram
\beq
{i}\Pi _1^\sigma(p) =  \int\frac{d^dk}{(2\pi)^d}\left[
-8 \frac{[(k\cdot p)^2 - k^2 p^2]^2}{k^4 (k+p)^4} 
+2 \frac{k^2 p^2+2(k\cdot p)^2 }{k^4}\right].
\eeq
In the $\sigma$ basis both diagrams are IR-convergent thanks to the numerators:
the bubble numerator vanishes for $k=0$ and for $k=-p$;
the seagull numerator vanishes for $k=0$.
The seagull, which carries no $\ln p$ dependence, vanishes in dimensional regularization.
The one loop result is 
\beq\frac{\Pi _1^\sigma(p)}{\Pi _0^\sigma(p)}\simeq
\frac{5 f_0^2}{12(4\pi)^2} \left[\frac{1}{\epsilon} +\ln \bigg(-\frac{\mub^2}{p^2}\bigg) + \frac35\right].
\label{eq:Pi1sigma}
\eeq
Once again, the UV-divergent part of eq.\eq{Pi1sigma} gives the usual UV $\beta$ function 
$\beta_{f_0^2} =5 f_0^4/6(4\pi)^2$, in agreement with eq.\eq{RGG}.
This same $1/\epsilon$ counter-term renormalises the 3-point and 4-point functions at one loop.
This time the quadratic action $\Pi _1^\sigma(p)$ contains no anomalous $\ln p$ dependence, in disagreement with eq.\eq{RGE4dp}.
We computed the same theory using two different field parameterisations, $h$ and $\sigma$:
this result confirms that (unlike scattering amplitudes and the usual UV RGE),
the dependence of the propagator on the off-shell momentum $p$ is unphysical, being affected by field re-parameterisations.


\subsection{Spin-0 sector in terms of the scale-factor $\Omega$}\label{eq:free}
The third alternative parameterisation of the graviton trace in terms of $\Omega$ in eq.\eq{g2trace}
gives a simpler action
\beq \label{eq:LagOmega}
\Lag_\Omega =  \frac{6}{f_0^2}\left[
\frac{ (\partial^2\Omega)^2}{\Omega^2}+ M_0^2\Omega \partial^2 \Omega\right]
-V \Omega^4.\eeq
It can be perturbatively computed by expanding $\Omega = 1 +  \omega$ in small fluctuations $\omega \ll 1$.
Tree-level amplitudes again vanish on-shell at $p_i^2=0$. 
The $\Omega$ basis makes this property manifest at the diagrammatic level:
now each tree diagram individually vanishes, since any vertex vanishes unless at least two particles have $p_i^2\neq 0$.
The general understanding of eq.\eq{Anselmi} for the vanishing of tree scattering amplitudes among the massless mode
gets now simplified noticing that the Einstein $R$ term just gives the kinetic term for a free $\Omega$ scalar,
and that the theory with the $R^2$ term is obtained from it by performing a field redefinition
$\Omega \to \Omega + \Omega^{-2} \partial^2\Omega/M_0^2 + {\cal O}(1/M_0^4)$, valid only locally~\cite{Marcus:1984ei,hep-th/0605205,1506.04589}.
On-shell tree amplitudes are the same as in the previous $h$ and $\sigma$ parameterizations, eq.~(\ref{sys:Amplitudes}).
In the $\Omega$ basis all diagrams contribute at high energy, but none produces terms that grow with higher powers of energy.
This suggests that the $\Omega$ basis partially avoids the unnecessary off-shell complications seen in the other parameterisations.

The tree level 2-particle action is $\Pi_0^\Omega  (p) = 12 p^2(p^2-M_0^2)/f_0^2$. 
The one loop correction in the limit $M_0=0$ from the bubble and the seagull diagrams is
\beq {i}\Pi_1^\Omega (p) =   \int\frac{d^dk}{(2\pi)^d}\left[-2\frac{[k^4+p^4+2(p^2+k^2)(k\cdot p)+3p^2 k^2 ]^2}{k^4(k+p)^4}+
3\frac{k^4+p^4+4k^2p^2}{k^4}\right].
\eeq
Both diagrams are IR and UV divergent.
Restoring the $M_0$ mass to control IR divergences
and computing in the off-shell limit $p\gg M_0$ gives
\beq \frac{ \Pi_1^\Omega(p)}{ \Pi_0^\Omega(p) } \simeq \frac{5  f_0^2}{12(4\pi)^2} \left[\frac{1}{\epsilon} +\ln \frac{-\mub^2}{M_0^{2/5}p^{8/5}} +\frac{9}{5}\right].
\label{eq:Pi1omega}
\eeq
This correction to the propagator reproduces the same UV divergence as in eq.\eq{Pi1h} and thereby the same UV $\beta$ function for $f_0$.
Like in the $h$ and $\sigma$ bases, eq.\eq{Pi1omega} contains an IR divergence which results in a different coefficient for the $\ln p$ term.
However, the $\ln p$ coefficient in eq.\eq{Pi1omega} 
differs from the previous $h$-basis computation in eq.\eq{Pi1h} 
and thereby from the claim of~\cite{2403.02397} in eq.\eq{RGE4dp},
confirming that it's a parameterisation-dependent artefact.
Unlike in the $h$ and $\sigma$ parameterisations,
 the one-loop propagator now has a simple  Passarino-Veltman decomposition without terms apparently enhanced by powers of $p/M_0$:
\beq\label{eq:PiOmega1}
\Pi_1^\Omega (p)=
-4  p^4 B_0(p,0,M_0^2)+2 (M_0^2+2 p^2)^2 B_0(p,M_0^2,M_0^2)+
   (p^4-4 M_0^2 p^2+M_0^4) \frac{A_0(M_0^2)}{M_0^2}.
\eeq
The UV-divergent part of the correction to the propagator is
\beq\label{eq:Pi1Omega}
(4\pi)^2 \Pi^{\Omega}_1(p)_{\rm div}= \frac{5  f_0^2}{12\epsilon} \Pi_0^\Omega + \frac{ 9M_0^2p^2}{\epsilon}  +  \frac{ 3M_0^4}{\epsilon} \\ 
\eeq
having used that the divergent parts in $d=4-2\epsilon$ dimensions of the Passarino-Veltman functions 
are $B_0 \simeq 1/(4\pi)^2\epsilon$ and $A_0(M_0) \simeq M_0^2 B_0(0,M_0^2,M_0^2)$.
The UV-divergent parts of the one-loop corrections to the 3 and 4-point 1PI amplitudes $\Gamma_3$ and $\Gamma_4$ are
\begin{eqnsystem}{sys:StrumiaOmegaM0}
(4\pi)^2 \Gamma_{31}(p_1^2, p_2^2, p_3^2)_{\rm div}&=& \frac{5 f_0^2 }{12\epsilon}  \Gamma_{30} +  \frac{6i M_0^4}{\epsilon},\qquad
 \Gamma_{30}=-i\frac{24}{f_0^2} (p_1^2p_2^2 + p_2^2 p_3^2 + p_3^2 p_1^2) ,
 \\
 (4\pi)^2 \Gamma_{41}(p_1^2,p_2^2,p_3^2,p_4^2)_{\rm div}&=& \frac{5f_0^2 }{12\epsilon} \Gamma_{40}+  \frac{6i M_0^4}{\epsilon},
 \qquad   \Gamma_{40} =+i \frac{72}{f_0^2} (p_1^2 p_2^2 + \hbox{5~perms}),
\end{eqnsystem}
where $\Gamma_{30}$ and $\Gamma_{40}$ are the tree amplitudes.
We here used the non-canonical basis, where $\omega$ needs no wave-function renormalization and 
the theory is rendered finite by the renormalisation of $f_0, M_0$ and $V$. 
The terms proportional to $M_0^4$ are all cancelled by the $V$ counter-term.
The terms proportional to 4 powers of momenta are all cancelled by the $f_0$ renormalisation.
The second term in eq.\eq{Pi1Omega} is cancelled by a $M_0$ counter-term.
A shift of the $\Omega$ field gives a redundant combination of the previous renormalisations, as it describes a renormalisation of the unit of mass.

\subsubsection*{Physical amplitudes}
The greater simplicity of the $\Omega$ parameterisation allows to compute extra quantities, 
allowing us to test whether they receive IR enhancements.
One such physical quantity affected by off-shell propagators is the 
\beq \omega(p_1)\omega(p_2)\to\omega(q_1)\omega(q_2)\eeq
scattering amplitude $\mathscr{A}_{p_1 p_2\to q_1 q_2}/N$ computed by summing amputated reducible diagrams
at generic $s=(p_1+p_2)^2$, $t=(p_1-q_1)^2$, $u=(p_1-q_2)^2$ and at on-shell external momenta $p_i^2,q_i^2 = \{0, M_0^2\}$.
The LSZ normalization factor $N$ makes the external particles canonical.
It is the same at both poles, equal to $N=[12 M_0^2/f_0^2]^2 $ at tree level, and to
$N=[12 {M}_{0\rm pole}^2/{f}_0^2(M_0)]^2$ at loop level, where ${M}_{0\rm pole}$ is the renormalized pole mass,
given by $M_0$ plus quantum corrections,  and ${f}_0(M_0)$ is the $f_0$ coupling renormalized at $M_0$.
The $2\to 2$ scattering amplitude is obtained by combining the 1PI correction to the 4-point function $\Gamma_4$
with the correction to the 3-point function $\Gamma_3$ and  with the self-energy $\Pi$.
At tree level, the massless state scatters exclusively into massive states. Consequently, only $\Gamma_4$ contributes to the simplest
one-loop scattering amplitude $\mathscr{A}$ among massless modes, $p_{1,2}^2=q_{1,2}^2=0$,
given that the other contributions get multiplied by the vanishing tree amplitude. The one loop result is
\begin{eqnarray}\label{eq:A1loop}
\frac{\mathscr{A}_{00\to 00}}{N} &=&-i\frac{f_0^4}{72}\bigg[[B_0(s)+B_0(t)+B_0(u)]+\\
&&+ 8M_0^2 [C_0(s)+C_0(t)+C_0(u)]+
8M_0^4[D_0(s,t)+D_0(s,u)+D_0(t,u)\bigg]\nonumber
\end{eqnarray}
where all Passarino-Vetlman $B_0$, $C_0$ and $D_0$ functions involve massive $M_0$ propagators only,
so we omitted the masses.
The $C_0$ and $D_0$ functions are finite and are negligible in the high-energy limit $s,t,u\gg M_0^2$, so we omit their standard definition.
The UV-divergent $B_0$ terms are renormalised by $V$.


\smallskip

We next consider the $00 \to M_0 M_0$ scattering amplitude  for the production of two massive modes, thereby setting $p_i^2=0$ and $q_i^2=M_0^2$.
It is obtained by combining  the correction to the off-shell propagator in eq.\eq{PiOmega1} with the following off-shell corrections to 1PI cubics $\Gamma_3$
and with the on-shell 1PI quartic $\Gamma_4$:
\begin{eqnsystem}{sys:VittiOmegaM0}
\Gamma_{31}(0,0,s) &\simeq&4 i M_0^2 s [B_0(0)-B_0(s) ],\\
 \Gamma_{31}(0,M_0^2,s) &\simeq& 10 if_0^3M_0^2 s B_0(0),\\
 \Gamma_{41}(0,0,M_0^2 , M_0^2) &\simeq& 2 i f_0^4 M_0^4[2B_0(s)+B_0(t)+B_0(u)-22 B_0(0)],
\end{eqnsystem}
where we reported the high-energy limit, $s,t,u\gg M_0^2$. The amplitude at high-energy is:
\beq \label{eq:A00M0M0Omega}
\frac{\mathscr{A}_{00\to M_0 M_0}}{N} \stackrel{s,t,u\gg M_0^2}\simeq 
i\frac{{f}_0^4(M_0)}{6f_0^2}\bigg[1 + \frac{f_0^2}{12} (22B_0(0) - 6 B_0(s) - 7 B_0(t) - 7 B_0(u) )\bigg].
\eeq
Subtracting the $V$ counter-term it becomes
\beq \label{eq:Asubren}
\frac{\mathscr{A}_{00\to M_0 M_0} - \mathscr{A}_{00\to 0 0}}{N} \stackrel{s,t,u\gg M_0^2}\simeq 
i\frac{{f}_0^4(M_0)}{6f_0^2}\bigg[1 + \frac{5f_0^2}{12} \frac{22B_0(0) -5 B_0(s) - 6 B_0(t) - 6 B_0(u)}{5}\bigg].
\eeq
where $B_0(s) \simeq (1/\epsilon +\ln \mub^2/s)/(4\pi)^2$ at $s\gg M_0^2$.
This amplitude is renormalized by the $f_0$ counter-term.
One can note that, quite unusually, the high-energy limit is not approximated by the tree level term with $f_0$ renormalised at high energy.
In the next section we use an equivalent two-derivative action  to reproduce this unusual behaviour  
and to demonstrate that it is process-dependent,
as expected for IR enhancements. 
We do not if/how such effects can be resummed, like the different IR enhancements that arise in ordinary QFT.
Practically speaking, though, resumming the loop contribution is unlikely to be necessary.
Indeed, related IR  enhancements appearing at tree level indicate that
in quadratic gravity scatterings never reach an energy so high~\cite{1808.07883} that these large process-dependent loop logarithms  invalidate perturbation theory.
This is illustrated, in more general terms, in the concluding section.

These one-loop amplitudes have been obtained with the help of FeynCalc and PackageX~\cite{Mertig:1990an,1503.01469,1612.00009,2312.14089}.


\subsection{Spin-0 sector in the 2-derivative  basis}\label{sec:22}
One more form of the theory in eq.\eq{LagOmega} 
is obtained by rewriting $\Lag_\Omega$ in terms of two 2-derivative fields
$\Omega$ and $\Omega'$ as
\beq\label{eq:LagOmegaLR}
\Lag_{\Omega\Omega'} =  -2\Omega' \partial^2\Omega -\frac{f_0^2}{6}\Omega^2\Omega'^2 +\frac{6 M_0^2}{f_0^2} \Omega\partial^2\Omega
- V \Omega^4.\eeq
Integrating out the auxiliary field $\Omega' = -6\partial^2\Omega/f_0^2\Omega^2$ gives back eq.\eq{LagOmega}.
The action in eq.\eq{LagOmegaLR} can be expanded around $\med{\Omega}=1$ (a different value just rescales the unit of mass) and
diagonalised by a field transformation that gets singular for $M_0\to 0$,
\beq \label{eq:SingTr}
\Omega = 1 + \frac{ f_0  (\alpha+\beta)}{\sqrt{12}M_0},\qquad
\Omega'=  \frac{\sqrt{3} M_0}{f_0} \beta.\eeq
The action becomes~\cite{1705.03896}
\beq
\Lag_{\alpha\beta} = -  \frac{(\partial \alpha)^2}{2} + \frac{(\partial\beta)^2-  M_0^2\beta^2}{2} -\frac{f_0^2}{24} \beta^2 (\alpha+\beta) \bigg(\alpha+\beta+ 4\sqrt{3}\frac{M_0}{f_0}\bigg) - V \Omega^4. \label{eq:abLag}\eeq
The field $\alpha$ is the massless Einstein ghost, while $\beta$ is the massive scalar.
The propagator of their $\alpha+\beta$ combination exhibits a ghost-like suppression at $p\gg M_0$.
Scattering amplitudes involving $\alpha$ only (such as $\alpha \alpha \to\alpha\alpha,\alpha\alpha\alpha$)
trivially vanish at tree level for $V=0$: no diagrams exist since vertices involve at least two $\beta$.
The $2\to 2$ scattering amplitudes among $\alpha$ and $\beta$ at tree level reproduce, up to an irrelevant overall sign, 
the corresponding scattering amplitudes among the states with masses $0$ and $M_0$ in previous formulations. 
In the high-energy limit, these amplitudes are predominantly governed by quartic couplings, which yield the constant terms in 
eq.~(\ref{sys:Amplitudes}). Diagrams involving cubic couplings contribute additional terms that include propagators. 
This behaviour arises because $\Lag_{\alpha\beta}$  features non-derivative potential interactions. 
By avoiding unnecessary off-shell complications, the $\alpha\beta$ basis 
provides a clearer framework for elucidating the origin of infrared-enhanced logarithmic terms.

\smallskip

For the same reason, the off-shell structure of $\Lag_{\alpha\beta}$
is different and simpler from that of the four-derivative $\Lag_\Omega$:
the UV-divergent corrections are momentum-independent.
We verified at one loop level that $\Lag_{\alpha\beta}$ is renormalizable, despite the apparently arbitrary coefficients in the potential.
At one loop the theory is renormalized by UV-divergent corrections to $f_0, M_0, V$, by a $\beta \to \beta + \delta\beta$ shift,  by a SO(1,1) transformation,
\beq \alpha \to\alpha+ \theta_{\alpha\beta}\beta,\qquad \beta\to \beta+ \theta_{\alpha\beta}\alpha.\eeq
A shift $\alpha \to \alpha+\delta\alpha$ is equivalent to a combination of the transformations above.
Wave-function corrections are UV finite.
We compute the one-loop tadpoles for $\alpha$ and $\beta$  
\beq T_\beta=3 T_\alpha=\sqrt{3}f_0 M_0 A(M_0^2)/2\eeq
which fix $V$ and $\delta\beta$, non-vanishing in view of the factor 3.
We then renormalise them away, dropping diagrams with $\alpha$ and $\beta$ tadpoles and including the counter-terms into the effective action.

The key non-trivial feature responsible for the emergence of unusual logarithmic terms is that  $\theta_{\alpha\beta}$ affects the quartic couplings, 
despite being determined by corrections to dimensionful terms: the tadpoles $T_{\alpha,\beta}$ and the mass mixing term $\Pi_{\alpha\beta}(0)$.
The only UV-divergent correction to the propagators are
\beq \Pi_{\alpha\alpha}(0)= \frac{f_0^2}{4}A(M_0^2),
\qquad \Pi_{\alpha\beta}(0)=3\Pi_{\alpha\alpha}(0),\qquad
\Pi_{\beta\beta}(0)= \frac{20}{3}\Pi_{\alpha\alpha}(0).\eeq
No wave-function renormalisations or momentum-dependent ($p$-running) contributions appear in the off-shell propagators.
Nevertheless, the one loop propagator for the $\Omega$ combination in eq.\eq{SingTr} reproduces the $\Omega$-basis one loop propagator
(up to a $p^2 B_0(0)$ term that renormalises\footnote{Since our scalars describe the graviton trace, different field variables
correspond to different units of mass, explaining why the dimensionful $M_0$ parameter receives
apparently different renormalisations: they are actually  physically identical, but expressed in different units
across different parameterisations in field space. }  $M_0^2$).
The 1PI cubic amplitudes receive the following UV-divergent corrections:
\beq \begin{array}{ll}\displaystyle
\Gamma_{\alpha\alpha\alpha}^{\rm div}=-\frac{f_0^3 M_0}{4\sqrt{3}} \frac{1}{(4\pi)^2\epsilon}, \qquad & \displaystyle
\Gamma^{\rm div}_{\alpha\alpha\beta}=3\Gamma_{\alpha\alpha\alpha}^{\rm div} ,\\
\displaystyle
\Gamma^{\rm div}_{\alpha\beta\beta}=-i\frac{f_0 M_0}{\sqrt{3}(4\pi)^2\epsilon} +  \frac{20}{3}\Gamma_{\alpha\alpha\alpha}^{\rm div} ,& \displaystyle
 \Gamma^{\rm div}_{\beta\beta\beta}=-i\frac{\sqrt{3}f_0 M_0}{(4\pi)^2\epsilon}  + 12\Gamma_{\alpha\alpha\alpha}^{\rm div} .
 \end{array}
 \eeq
The $2\to 2$ 1PI effective vertices receive one loop corrections.
We present both the ultraviolet (UV)-divergent terms and the finite terms that dominate at high energies.
In the current basis, these corrections arise exclusively from diagrams involving only quartic coupling:
\begin{eqnsystem}{sys:Aalphabeta}
\Gamma_{\alpha\alpha\to\alpha\alpha}&\simeq&i\frac{f_0^4}{72}[B_0(s)+B_0(t)+B_0(u)] ,\\
\Gamma_{\alpha\alpha\to\alpha\beta} &\simeq&3\Gamma_{\alpha\alpha\to\alpha\alpha}, \\
\Gamma_{\alpha\alpha\to\beta\beta}&\simeq&- i\frac{f_0^2}{6} +i\frac{f_0^4}{72}[6B_0(s)+7B_0(t)+7B_0(u)],\\ 
\Gamma_{\alpha\beta\to\beta\beta} &\simeq& -i\frac{f_0^2}{2}+12\Gamma_{\alpha\alpha\to\alpha\alpha},\\
\Gamma_{\beta\beta\to\beta\beta}&\simeq& -i f_0^2 +19\Gamma_{\alpha\alpha\to\alpha\alpha}.
\end{eqnsystem}
Concerning the divergent parts,
$\Gamma_{\alpha\alpha\to\alpha\alpha}$ is renormalized by $V$ only, that contributes to all quartics in the same way.
$\Gamma_{\alpha\alpha\to\alpha\beta} $ also by the SO(1,1) rotation $\theta_{\alpha\beta}$.
$\Gamma_{\alpha\alpha\to\beta\beta}$ and the others also by $f_0$.

In the $\alpha\beta$ theory, the high-energy limits of the scattering amplitudes are simply dominated by the 1PI diagrams,
as the extra corrections to cubics and propagators (including the Lehmann-Symanzik-Zimmermann factors) are suppressed at high energy.
In particular, $\mathscr{A}_{\alpha\alpha\to\alpha\alpha}\simeq\Gamma_{\alpha\alpha\to\alpha\alpha}$ agrees with the $\Omega$ basis 
result of  eq.\eq{A1loop},
and $\mathscr{A}_{\alpha\alpha\to\beta\beta}\simeq \Gamma_{\alpha\alpha\to\beta\beta}$ agrees with the $\Omega$ basis result of eq.\eq{A00M0M0Omega},
provided they are evaluated at common values of the free parameters following the renormalisation procedures.
So eq.s~(\ref{sys:Aalphabeta}) exhibit the unusual feature: 
physical amplitudes at high-energy receive log-enhanced corrections with process-dependent
coefficients different from the RG running of the $f_0$ coupling that appears in the tree amplitudes.
For example the renormalized value of an amplitude combination not affected by the $V$ counter-term is
\beq \mathscr{A}_{\alpha\alpha\to\beta\beta}-\mathscr{A}_{\alpha\alpha\to\alpha\alpha}
\stackrel{s,t,u\gg M_0^2}\simeq -i \frac{f_0^2(\mub)}{6} \bigg[1 - \frac{5 f_0^2}{12(4\pi)^2} \ln\frac{ \mub^{2} s (tu)^{6/5} }{M_0^{44/5}}\bigg]\eeq
in agreement with eq.\eq{Asubren}.
Unlike in a normal QFT, the limit $M_0\to 0$ is not naive, as
cross sections at one loop are not roughly given by their tree level expressions,
inserting couplings renormalized at the energy of the process, $\bar\mu^2\sim s\sim t\sim u$.
Such unusual logarithms are compatible with renormalizability because of the unusual origin of the $\theta_{\alpha\beta}$ renormalisation.

\subsection{Why the limit   $M_0\to 0$ is singular}
We finally discuss what leads to log-enhanced corrections that are unusual in QFT.

In conventional QFT, the high-energy behaviour is obtained by naively taking the limit $M_0 \to 0$ for a mass term.
 However, in theories with ghost fields, this limit is singular due to IR divergences.

The determinant $1/p^2 (p^2- M_0^2)$ of the propagator matrix of $\Omega,\Omega'$ in eq.\eq{LagOmegaLR} 
reproduces the propagator of the  equivalent 4-derivative theory,
and is negligibly affected by $M_0$ at large energy $p\gg M_0$.
However, the eigenvectors of this matrix are significantly influenced by $M_0$. 
Indeed, the transformation in eq.\eq{SingTr} diagonalises the kinetic and mass term through a SO(1,1) rotation in field space which becomes singular as $M_0\to 0$.
The boost parameter of a SO(1,1) rotation can become arbitrarily large, unlike the angle of a SO(2) rotation that diagonalises a theory without ghosts.
This feature, implicit in the single-field formulation with four derivatives, becomes explicit when the same theory is
reformulated, as in  eq.\eq{abLag}, in terms of a normal scalar $\beta$ with mass $M_0$ interacting with a ghost $\alpha$.
Consequently, a small $M_0$ not only regularizes IR divergences but also plays a critical role in defining the scattering states.

\medskip
The 4-derivative actions for the spin 0 sector of quadratic gravity inherit renormalizability from the full theory
(where renormalizability  follows from simple power-counting, as the couplings $f_{0,2}$ are dimension-less).
The ghost-like nature of $\alpha$ is crucial for renormalizability:  if the sign of the $\alpha$ kinetic term were flipped, 
$\Lag_{\alpha\beta}$ would no longer be renormalizable, 
and the unusual high-energy logarithms in the amplitudes of eq.~(\ref{sys:Aalphabeta})
would be trivially understood as multiple quartics with different $\beta$ functions.
In the ghost-like theory there is a unique quartic $f_0$, and
the unusual logarithms arise because the $M_0$ term is renormalized by a ${\rm SO}(1,1)$ rotation that affects the quartics.

\smallskip

Since the SO(1,1) rotation in eq.~(\ref{eq:SingTr}) gets singular for $M_0\to 0$, the term proportional to $M_0^2$ in $\Lag_{\Omega\Omega'} $, eq.~(\ref{eq:LagOmegaLR}), 
gives a contribution to the kinetic terms in $\Lag_{\alpha\beta}$, eq.~(\ref{eq:abLag}), even in the $M_0\to 0$ limit. 
So, the $M_0\to 0$ limit taken in~(\ref{eq:abLag}) is not eq.~(\ref{eq:LagOmegaLR}) with $M_0$  set to 0,
\beq\label{eq:LagOmegaLR0}
\Lag_{\Omega\Omega'} (M_0 = 0)=  -2\Omega' \partial^2\Omega -\frac{f_0^2}{6}\Omega^2\Omega'^2
 .\eeq
To confirm that this subtlety is the source of the unusual high-energy logarithmic corrections, we show that these corrections  
do not arise from the different theory in eq.\eq{LagOmegaLR0}.
This theory is invariant under $\Omega\to \lambda \Omega$, $\Omega'\to \Omega'/\lambda$.
So the two scalars form a SO(1,1) doublet
 \beq \vec\Omega = (\Omega_+,\Omega_-)= (m \Omega+\Omega'/m, \Omega'/m-m\Omega),\eeq
 where $m$ is an arbitrary mass.
 The field-space metric is $\diag(1,-1)$ such that
\beq\Lag_{\Omega\Omega'} (M_0 = 0)=  \frac{(\partial\Omega_+)^2}{2} - \frac{(\partial\Omega_-)^2}{2} -  \frac{f_0^2}{96} (\Omega_+^2 - \Omega_-^2)^2=
 \frac{(\partial\vec\Omega)^2}{2} - \frac{f_0^2}{96}\vec\Omega^4.\label{VecOmegaLag}\eeq
 The apparently `soft' term proportional to $M_0^2$ in eq.~(\ref{eq:LagOmegaLR}), 
 despite being suppressed by a possibly small mass $M_0$,  breaks SO(1,1) in a hard way, giving a kinetic term
 \beq \label{eq:extraM0} 
\frac{6 M_0^2}{f_0^2} \Omega\partial^2\Omega  = \frac{3M_0^2}{2f_0^2 m^2} (\Omega_+ - \Omega_-)\partial^2 (\Omega_+ - \Omega_-).\eeq
If this term is ignored, the SO(1,1)-invariant theory can be simply computed by expanding around $\vec\Omega=0$, finding that it's not the same theory.
The only correction to the propagator is a vanishing seagull $A_0(M_0)\to A_0(0)=0$. 
The  correction to the quartics reproduces the RGE for $f_0$, as this $\beta$ function is insensitive to the precise definition of the IR states.
The two independent scattering amplitudes at one-loop  
\begin{eqnsystem}{sys:A+-} 
\mathscr{A}_{\Omega_-\Omega_-\to \Omega_-\Omega_- }&=&+ i \frac{f_0^2}{4}\left[1-\frac{5f_0^2}{12} \frac{B_0(s)+B_0(t)+B_0(u)}{3}\right],\\
\mathscr{A}_{ \Omega_-\Omega_-\to \Omega_+\Omega_+}&=&- i \frac{f_0^2}{12}\left[1-\frac{5f_0^2}{12}  \frac{3B_0(s)+B_0(t)+B_0(u)}{5}\right],
\end{eqnsystem}
are both renormalized by $f_0$ in the usual way, exhibiting no unusual logarithmic enhancements.
So, when combining them to compute any $2\to 2$ scattering amplitude,
they are all approximated by their tree level value, appropriately renormalized at the scale of the scattering process.
These amplitudes differ from the amplitudes of the $\Lag_{\alpha\beta}$ theory, even at tree level.

\section{Discussions and conclusions}\label{sec:concl}
This paper is technically intricate, involving several subtleties and differences with earlier literature.
We conclude by setting  these complications aside to present a summary of our overall understanding.

\medskip

We clarified a  novel QFT feature that appears in theories in $d=3+1$ dimensions with 4 derivatives ---
or equivalently, those involving ghost states.
At the loop level, logarithmically enhanced corrections modify the high-energy behaviour of physical cross sections. 
Unlike in conventional QFT, these cross sections are not accurately described by their tree-level forms with running couplings simply renormalized at high energies. 
This phenomenon arises in a theory of particular physical interest: quadratic gravity. While our detailed calculations focused on this theory, 
the result is more general and extends to other four-derivative theories, such as higher-derivative gauge or scalar models.

We find that these unusual logarithms appear as process-dependent IR enhancements.
They cannot be reabsorbed in modified $\beta$ functions for the couplings, that could make quadratic gravity asymptotically free.
Instead, these IR-enhanced logarithms could perhaps be resummed by interpreting them as an `IR dressing' of the scattering states.

Indeed, the origin of these enhanced logarithms lies in the structure of four-derivative fields, which propagate two distinct mass shells
--- one associated with a normal mode (positive kinetic energy) and one with a ghost (negative kinetic energy).
Physical scattering amplitudes, independent of gauge choice and field parameterisation,
are obtained from residuals at both mass shells of the off-shell amplitudes of the 4-derivative field.
As a result, small mass terms that kinematically differentiate the two states defining them play an unusually relevant role
at energies much larger than the masses, resulting in infra-red-enhanced logarithms.

This behaviour becomes more transparent when reformulating one four-derivative field as a pair of two-derivative fields: one with positive and one with negative kinetic energy. In the limit where their masses become degenerate, the theory exhibits a singular behaviour characterized by a large SO(1,1) boost in field space,
needed to diagonalise the mass and kinetic terms.
In this way, super-renormalizable mass terms can influence dimension-less interactions even at high energies.

It is important to note, however, that in quadratic gravity these unusual logarithms appear in the regime of sub-Planckian energies and masses where the theory is perturbatively calculable
(its couplings $f_{0,2}\sim M_{0,2}/\bp$ are Planck-suppressed) and do not ruin its calculability.
Indeed, tree-level IR enhancements become large at Planckian energies,
and screen
trans-Planckian physics: trans-Planckian collisions are inhibited because energy is radiated away through the emission of many soft gravitons~\cite{1808.07883}. As a result, the process-dependent enhanced logarithms 
are never too big to invalidate perturbation theory because, effectively, the energies can never be trans-Planckian.

QFT in $d=1+1$ dimensions too are  expected to feature unusual IR-enhanced logarithms.
Such theories do not have ghosts, calling for a different interpretation.

\medskip

As a byproduct, we partially computed the effective field theory of quadratic gravity, obtained by integrating out its massive modes.
At tree level, there is no contribution ---  implying that the ghost state remains invisible to EFT positivity bounds~\cite{2103.12728,2201.06602}.
A non-vanishing effect is expected to arise at one loop level, where we only computed the simpler spin 0 sector.
Adding matter in addition to pure gravity, the ghost becomes visible at tree level~\cite{2110.02246}.


\small

\subsubsection*{Acknowledgments}
We thank L.~Buoninfante, D.~Buccio, J.F.\ Donoghue, R.~Percacci and O.~Zanusso  for useful discussions.

\footnotesize

\begingroup
\renewcommand{\baselinestretch}{0.95}\selectfont

\endgroup


\begin{thebibliography}{nnn}\bibitem{2307.00055}
\article[2307.00055]{D. Buccio, J.F. Donoghue, R. Percacci}{Phys.Rev.D}{109}{045008}{2024}
{\href{https://doi.org/10.1103/PhysRevD.109.045008}{Amplitudes and renormalization group techniques: A case study}}.


\bibitem{2403.02397}
\article[2403.02397]{D. Buccio, J.F. Donoghue, G. Menezes, R. Percacci}{Phys.Rev.Lett.}{133}{021604}{2024}
{\href{https://doi.org/10.1103/PhysRevLett.133.021604}{Physical Running of Couplings in Quadratic Gravity}}.


\bibitem{2408.13142}
\heparticle[2408.13142]{D. Buccio, J.F. Donoghue, G. Menezes, R. Percacci}{Renormalization and running in the 2D $CP(1)$ model}.


\bibitem{Avramidi:1985ki}
\article{I.G. Avramidi, A.O. Barvinsky}{Phys.Lett.B}{159}{269}{1985}
{\href{https://doi.org/10.1016/0370-2693(85)90248-5}{Asymptotic freedom in higher derivative quantum gravity}}.


\bibitem{Julve:1978xn}
\article{J. Julve, M. Tonin}{Nuovo Cim.B}{46}{137}{1978}
{\href{https://doi.org/10.1007/BF02748637}{Quantum Gravity with Higher Derivative Terms}}.


\bibitem{Fradkin:1981iu}
\article{E.S. Fradkin, A.A. Tseytlin}{Nucl.Phys.B}{201}{469}{1982}
{\href{https://doi.org/10.1016/0550-3213(82)90444-8}{Renormalizable asymptotically free quantum theory of gravity}}.


\bibitem{1403.4226}
\article[1403.4226]{A. Salvio, A. Strumia}{JHEP}{06}{080}{2014}
{\href{https://doi.org/10.1007/JHEP06(2014)080}{Agravity}}.


\bibitem{1808.07883}
\article[1808.07883]{A. Salvio, A. Strumia, H. Veerm{\" a}e}{Eur.Phys.J.C}{78}{842}{2018}
{\href{https://doi.org/10.1140/epjc/s10052-018-6311-1}{New infra-red enhancements in 4-derivative gravity}}.


\bibitem{Nielsen}
K. Nielsen, Nucl. Phys. B 101 (1975) 173;
R. Fukuda and T. Kugo, Phys. Rev. D13 (1976) 3469;
  I.~J.~R.~Aitchison and C.~M.~Fraser,
  Annals Phys.\  {156} (1984) 1;
  D.~Binosi, J.~Papavassiliou and A.~Pilaftsis,
  Phys.\ Rev.\ D {71} (2005) 085007
  [hep-ph/0501259].


\bibitem{1307.3536}
\article[1307.3536]{D. Buttazzo, G. Degrassi, P.P. Giardino, G.F. Giudice, F. Sala, A. Salvio, A. Strumia}{JHEP}{12}{089}{2013}
{\href{https://doi.org/10.1007/JHEP12(2013)089}{Investigating the near-criticality of the Higgs boson}}.


\bibitem{Chisholm:1961tha}
\article{J.S.R. Chisholm}{Nucl.Phys.}{26}{469}{1961}
{\href{https://doi.org/10.1016/0029-5582(61)90106-7}{Change of variables in quantum field theories}}.


\bibitem{Kamefuchi:1961sb}
\article{S. Kamefuchi, L. O'Raifeartaigh, A. Salam}{Nucl.Phys.}{28}{529}{1961}
{\href{https://doi.org/10.1016/0029-5582(61)90056-6}{Change of variables and equivalence theorems in quantum field theories}}.


\bibitem{tHooft:1973wag}
\article{G. 't Hooft, M.J.G. Veltman}{NATO Sci.Ser.B}{4}{177}{1974}
{\href{https://doi.org/10.1007/978-1-4684-2826-1_5}{Diagrammar}}.


\bibitem{1205.3279}
\article[1205.3279]{D. Anselmi}{Eur.Phys.J.C}{73}{2338}{2013}
{\href{https://doi.org/10.1140/epjc/s10052-013-2338-5}{A General Field-Covariant Formulation Of Quantum Field Theory}}.


\bibitem{2105.11482}
\article[2105.11482]{A. Baldazzi, R.B.A. Zinati, K. Falls}{SciPost Phys.}{13}{085}{2022}
{\href{https://doi.org/10.21468/SciPostPhys.13.4.085}{Essential renormalisation group}}.


\bibitem{2312.06748}
\article[2312.06748]{T. Cohen, X. Lu, D. Sutherland}{JHEP}{06}{149}{2024}
{\href{https://doi.org/10.1007/JHEP06(2024)149}{On amplitudes and field redefinitions}}.


\bibitem{Marcus:1984ei}
\article{N. Marcus, A. Sagnotti}{Nucl.Phys.B}{256}{77}{1985}
{\href{https://doi.org/10.1016/0550-3213(85)90386-4}{The Ultraviolet Behavior of $N=4$ {Yang-Mills} and the Power Counting of Extended Superspace}}.


\bibitem{1605.03602}
\article[1605.03602]{R. Alonso, E.E. Jenkins, A.V. Manohar}{JHEP}{08}{101}{2016}
{\href{https://doi.org/10.1007/JHEP08(2016)101}{Geometry of the Scalar Sector}}.


\bibitem{Stelle:1976gc}
\article{K.S. Stelle}{Phys.Rev.D}{16}{953}{1977}
{\href{https://doi.org/10.1103/PhysRevD.16.953}{Renormalization of Higher Derivative Quantum Gravity}}.


\bibitem{hep-th/0412249}
\article[hep-th/0412249]{G. de Berredo-Peixoto, I.L. Shapiro}{Phys.Rev.D}{71}{064005}{2005}
{\href{https://doi.org/10.1103/PhysRevD.71.064005}{Higher derivative quantum gravity with Gauss-Bonnet term}}.


\bibitem{1308.3398}
\article[1308.3398]{N. Ohta, R. Percacci}{Class.Quant.Grav.}{31}{015024}{2014}
{\href{https://doi.org/10.1088/0264-9381/31/1/015024}{Higher Derivative Gravity and Asymptotic Safety in Diverse Dimensions}}.

\bibitem{2305.10591}
\article[2305.10591]{H.~Kawai, N.~Ohta}{Phys.Rev.D}{107}{126025}{2023}
{\href{https://doi.org/10.1103/PhysRevD.107.126025}{Wave function renormalization and flow of couplings in asymptotically safe quantum gravity}}.



\bibitem{hep-th/0605205}
\article[hep-th/0605205]{D. Anselmi}{JHEP}{01}{062}{2007}
{\href{https://doi.org/10.1088/1126-6708/2007/01/062}{Renormalization and causality violations in classical gravity coupled with quantum matter}}.


\bibitem{1506.04589}
\article[1506.04589]{P. Don{\` a}, S. Giaccari, L. Modesto, L. Rachwal, Y. Zhu}{JHEP}{08}{038}{2015}
{\href{https://doi.org/10.1007/JHEP08(2015)038}{Scattering amplitudes in super-renormalizable gravity}}.


\bibitem{2103.12728}
\article[2103.12728]{Z. Bern, D. Kosmopoulos, A. Zhiboedov}{J.Phys.A}{54}{344002}{2021}
{\href{https://doi.org/10.1088/1751-8121/ac0e51}{Gravitational effective field theory islands, low-spin dominance, and the four-graviton amplitude}}.


\bibitem{2201.06602}
\article[2201.06602]{S. Caron-Huot, Y.-Z. Li, J. Parra-Martinez, D. Simmons-Duffin}{JHEP}{05}{122}{2023}
{\href{https://doi.org/10.1007/JHEP05(2023)122}{Causality constraints on corrections to Einstein gravity}}.


\bibitem{2304.02550}
\article[2304.02550]{B. Bellazzini, G. Isabella, S. Ricossa, F. Riva}{Phys.Rev.D}{109}{024051}{2024}
{\href{https://doi.org/10.1103/PhysRevD.109.024051}{Massive gravity is not positive}}.


\bibitem{1705.03896}
\article[1705.03896]{A. Salvio, A. Strumia}{Eur.Phys.J.C}{78}{124}{2018}
{\href{https://doi.org/10.1140/epjc/s10052-018-5588-4}{Agravity up to infinite energy}}.


\bibitem{Salvio:2024joi}
\article[2404.08034]{A.~Salvio}{JCAP}{07}{092}{2024}
{\href{https://doi.org/10.1088/1475-7516/2024/07/092}{A non-perturbative and back\-ground-independent formulation of quadratic gravity}}.


\bibitem{Mertig:1990an}
\article{R. Mertig, M. Bohm, A. Denner}{Comput.Phys.Commun.}{64}{345}{1991}
{\href{https://doi.org/10.1016/0010-4655(91)90130-D}{FEYN CALC: Computer algebraic calculation of Feynman amplitudes}}.


\bibitem{1503.01469}
\article[1503.01469]{H.H. Patel}{Comput.Phys.Commun.}{197}{276}{2015}
{\href{https://doi.org/10.1016/j.cpc.2015.08.017}{Package-X: A Mathematica package for the analytic calculation of one-loop integrals}}.


\bibitem{1612.00009}
\article[1612.00009]{H.H. Patel}{Comput.Phys.Commun.}{218}{66}{2017}
{\href{https://doi.org/10.1016/j.cpc.2017.04.015}{Package-X 2.0: A Mathematica package for the analytic calculation of one-loop integrals}}.


\bibitem{2312.14089}
\article[2312.14089]{V. Shtabovenko, R. Mertig, F. Orellana}{Comput.Phys.Commun.}{306}{109357}{2025}
{\href{https://doi.org/10.1016/j.cpc.2024.109357}{FeynCalc 10: Do multiloop integrals dream of computer codes?}}.


\bibitem{2110.02246}
\article[2110.02246]{B. Holdom}{JHEP}{04}{133}{2022}
{\href{https://doi.org/10.1007/JHEP04(2022)133}{Photon-photon scattering from a UV-complete gravity QFT}}.


\end{thebibliography}
\end{document}